\documentclass[12pt,preprint]{aastex}
\newcommand{ \BLskip }{ } 
\newcommand{\etal }{et al.}
\def\lesssim{\mathrel{\hbox{\rlap{\hbox{\lower4pt\hbox{$\sim$}}}\hbox{$<$}}}}
\def\gtrsim{\mathrel{\hbox{\rlap{\hbox{\lower4pt\hbox{$\sim$}}}\hbox{$>$}}}}
\shorttitle{Disk Emergence and Planet Formation}
\shortauthors{Inutsuka et al. 2009}
\begin{document}
\title{Emergence of Protoplanetary Disks and 
       Successive Formation of Gaseous Planets by Gravitational Instability}
\author{Shu-ichiro Inutsuka\altaffilmark{1}, 
        Masahiro N. Machida\altaffilmark{2}, 
    and Tomoaki Matsumoto\altaffilmark{3}} 
\altaffiltext{1}{Department of Physics, Nagoya University, 
   Furo-cho, Chikusa-ku, Nagoya, Aichi 464-8602, Japan; 
   inutsuka@nagoya-u.jp}
\altaffiltext{2}{National Astronomical Observatory, Mitaka, 
                 Tokyo 181-8588, Japan; masahiro.machida@nao.ac.jp}
\altaffiltext{3}{Faculty of Humanity and Environment, Hosei University, 
                 Fujimi, Chiyoda-ku, Tokyo 102-8160, Japan; 
				   matsu@i.hosei.ac.jp}
\begin{abstract}
\BLskip
We use resistive magnetohydrodynamical simulations  
 with the nested grid technique  
 to study the formation of protoplanetary disks around 
 protostars from molecular cloud cores 
 that provide the realistic environments for planet formation. 
We find that gaseous planetary-mass objects are formed 
 in the early evolutionary phase 
 by gravitational instability in regions that are decoupled from 
 the magnetic field and surrounded by the injection points of 
 the magnetohydrodynamical outflows during the formation phase of 
 protoplanetary disks. 
Magnetic decoupling enables massive disks to form and 
 these are subject to gravitational instability, even at $\sim$ 10 AU. 
The frequent formation of planetary-mass objects in the disk suggests 
 the possibility of constructing a hybrid planet formation scenario,  
 where the rocky planets form later under the influence of 
 the giant planets in the protoplanetary disk. 
\end{abstract}
\BLskip
\section{Introduction}
Recent direct imaging of outer planets in extra-solar planetary systems 
 \citep{Kalas+2008,Marois+2008,Lagrange+2008,Greaves+2008,Thalmann+2009}
 provides a challenging question to the theory of planet formation: 
 how are giant planets formed in the distant regions far from 
 the central star?
Almost all the dynamical timescales for the various important processes 
 in planet formation essentially scale with the Kepler rotation timescale 
 \citep[e.g.,][]{KokuboIda2002}. 
In the standard core-accretion scenario,
 there are severe timescale constraints on the in situ formation of 
 Jovian planets, even at 5 AU from the central star 
 \citep[e.g.,][]{Pollack+1996,IkomaNakazawaEmori2000,
 HubickyjBodenheimerLissauer2005,HoriIkoma2008,Lissauer+2009,Machida+2010}. 
Consequently, the formation of planets at greater distances 
 (up to $\sim 10^2$ AU) seems to be impossible within an observationally 
 reasonable timescale in the core-accretion scenario of planet formation 
 \citep[e.g., ][]{IdaLin2004}. 
On the other hand, 
 the formation of giant planets 
 due to gravitational instability of massive protoplanetary disks 
 has been extensively investigated by various authors  
 \citep[e.g., see review by][]{Durisen+2007}. 
However, most analyses focused only on the evolution of 
 hypothetical disks 
 and cannot be applied directly to actual systems. 
The result of gravitational instability or absence of it does depend on 
 how the protoplanetary disks are formed, 
 but the formation of the protoplanetary disk is closely related to 
 the formation of the central star. 
In other words, the initial conditions of planet formation by gravitational 
 instability should be provided by the star formation process. 

The last decade has seen dramatic progress in our understanding of 
 protostar formation \citep[e.g.,][]{AndreBasuInutsuka2008} 
 and now provides us with the opportunity to study the formation phase 
 of protoplanetary disks. 
A highlight of the recent non-ideal magnetohydrodynamical (MHD) 
 calculations of protostellar collapse from molecular cloud cores 
 is the driving of outflows from the first cores and well-collimated fast 
 jets from the protostars; these may be regarded as proof of the 
 importance of various physical processes such as the ohmic dissipation of 
 magnetic fields due to the low degree of ionization 
 at higher density phases. 
In this Letter, 
 we study the formation phase of protoplanetary disks 
 in a self-consistent, non-ideal MHD system. 
Our protostellar collapse calculations start from a molecular cloud core, 
 include (practically) all the realistic physical processes, 
 and show how knowledge of the protostar formation process provides 
 convincing evidence for the formation of giant planets in the early phase. 

\section{Non-Ideal MHD Simulations}
The initial condition of our resistive MHD simulation 
 is the critical Bonner-Ebert Sphere 
 (an isothermal sphere at gravitational equilibrium)  
 with a temperature of 10 K and a radius of 4750 AU. 
In order to initiate gravitational collapse, we increase the density 
 uniformly by a factor of 3 to an initial central density 
 of $3~\times~10^6~{\rm cm}^3$. 
The mass is 1.6 times the solar mass. 
In a typical simulation, we use a uniform rotation of 
 the angular velocity of $\Omega_{\rm init}=1.1\times10^{-13}{\rm s}^{-1}$ 
 and a uniform magnetic field strength of 
 $B_{\rm init}=37~\mu$ G in the initial state.  
The initial ratios of thermal, rotational, and magnetic energy to 
 the negated gravitational energy are 
 0.3, 0.005, 0.014, respectively 
 \citep[cf., ][]{Machida+2008c}. 
We adopt the nested-grid scheme in order to increase the spatial resolution 
 of the central region \citep{Machida+2005}; 
 the number of nest-grids is typically 12 
 and each grid has 
 $n_x~\times~n_y~\times n_z = 128~\times~128\times~32$ cells 
 \citep{Machida+2006c}. 
As a result, spatial resolution of the innermost grid is 0.58AU 
 and that of the outermost grid is 1200AU. 

To describe a realistic evolution of the magnetic field in protostar
 formation, we should take into account the non-ideal MHD
 effects of weakly ionized molecular gas. 
In general, the ambipolar diffusion is important in the low-density
 phase, but it is slow and not critical in the dynamically 
 collapsing state, 
In the intermediate-density phase, 
 the Hall term effect can produce a modest effect 
 depending on the size distribution of dust grains
 \citep{WardleNg1999}. 
In contrast, ohmic dissipation dominates in the high-density phase, 
 and is shown to be the most efficient mechanism for 
 the dissipation of magnetic field in the magnetically supercritical 
 cloud core 
 \citep[e.g.,][]{NakanoNishiUmebayashi2002}. 
Therefore, we model the non-ideal effects of the magnetic field 
 by the effective resistivity in the induction equation.
We adopt the resistivity evolution of the fiducial model of 
   \citet{MachidaInutsukaMatsumoto2007} that corresponds to 
 the ionization equilibrium in standard molecular clouds. 
As the cloud core collapses, the degree of ionization decreases with 
 increasing density. 
The gas becomes magnetically decoupled at densities of around 
 $10^{10} {\rm cm}^{-3}$ but couples again 
 when the temperature exceeds about 1000 K. 
The adopted equation of state is the same as that used in 
   \citet{Machida+2009} 
 that follows the radiation hydrodynamical 
 calculations of protostellar collapse \citep{MasunagaInutsuka2000}. 

\section{Results}
 
\begin{figure}
 \includegraphics[width=170mm]{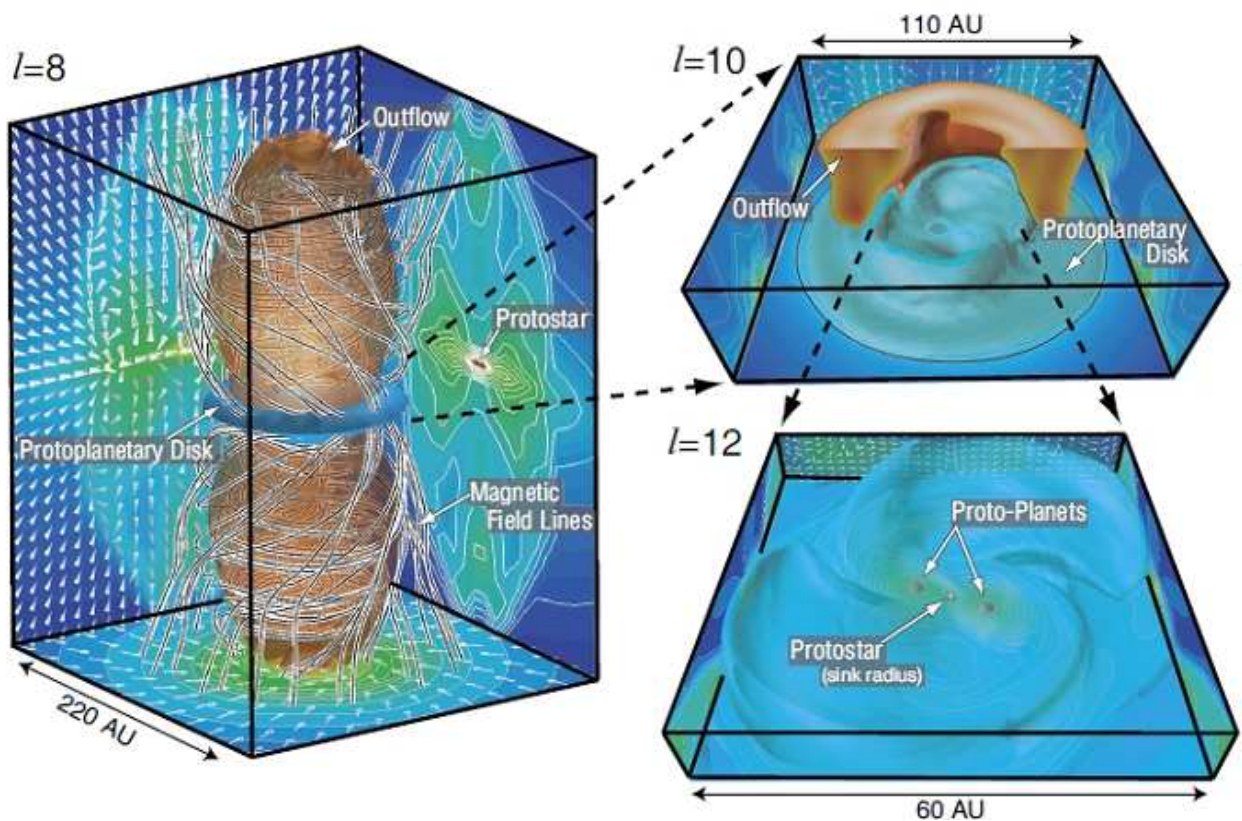}
\caption{
   \BLskip
   Bird's eye-view of the result of non-ideal MHD simulation with nested 
   grid technique, covering the evolution of the molecular cloud core 
   to the protostar. 
  The left panel shows the structure in the grid, level $l=8$, 
   where the high-density region 
   ($n=10^{10}{\rm cm}^{-3}$; blue isodensity surface)  
   and magnetic field lines are plotted. 
  Two cocoon-like structures (brown) above and below the flattened core 
   denote the zero-velocity surface inside of which  
   the gas is outflowing from the center. 
  The density contours (color and contour lines) and
    velocity vectors (thin white arrows) are projected in each wall surface.
 The right upper panel shows the structure in part of the 10th grid, 
  where we can clearly see the central cavity in the outflowing region. 
 The right lower panel (12th grid) shows 
  the protoplanetary disk in the formation phase, and 
  two newly formed planetary-mass objects in the disk. 
}
\label{fig:1}
\end{figure}
%
%
Figure 1 shows a typical ``bird's eye-view'' snapshot of our simulations. 
The timescale of the gravitational collapse (i.e., free-fall time) 
 is a decreasing function of density; 
 therefore, the dense central region shrinks faster than 
 the less-dense surrounding regions. 
This property of gravitational collapse almost always 
 leads to the successive decrease of mass inside 
 the faster shrinking region in a run-away manner, 
This gravitational ``run-away collapse'' 
 is decelerated by 
 a gradual increase in the temperature of the central region, 
 and eventually a quasi-steady object called ``the first core''
 is formed 
 \citep{Larson1969,WinklerNewman1980a,WinklerNewman1980b,
        MasunagaMiyamaInutsuka1998}. 
It consists mainly of hydrogen molecules and has a radial extent 
 exceeding 10 AU. 
The molecular gas surrounding the first core continues to accrete onto it, 
 resulting in a slow but monotonic increase in density and temperature 
 at the center of the first core. 
If the initial angular momentum of the molecular cloud core is 
 on the order of the value suggested by observation, 
 the resultant first core rotates significantly fast, 
 and its formation corresponds to 
 the onset of bipolar outflows driven by magnetic fields 
 \citep{Tomisaka2002, Machida+2006a, BanerjeePudritz2006, Machida+2008a, 
       HennebelleFromang2008}.
In Figure 1, the outflow regions are shown by 
 the two brown cocoon-like structure 
 that correspond to the zero vertical velocity surfaces ($v_z=0$): 
 gas inside the cocoons has a significant vertical velocity. 

Most of the angular momenta in gravitationally collapsing objects are 
 removed by the Maxwell stress of the field, which is called magnetic braking 
 \citep{MachidaInutsukaMatsumoto2007}. 
In addition the outflowing gas carries away angular momentum during this phase. 
When the central temperature becomes sufficiently high  
 ($\sim 2\times 10^3$K), the dissociation of hydrogen molecules 
 becomes significant, providing effective cooling that makes the core 
 gravitationally unstable, triggering ``the second collapse''
 \citep{MasunagaInutsuka2000, Machida+2006a, MachidaInutsukaMatsumoto2007, 
        Machida+2008a}. 

The upper right panel of Figure 1 shows an enlarged view of 
 the region inside the outflow-launching regions, 
 where the resistivity is so significant that 
 the magnetic field is decoupled from the gas. 
The outflow region envelopes  
 the ``dead zone'' for magnetic field 
 \citep{Machida+2008a}   
 where magnetic braking is not operating. 
Therefore, 
 the infalling gas in the first core maintains 
 angular momentum and reaches the radius of the centrifugal barrier 
 to form a circumstellar disk-like structure. 
The formation of this disk-like structure corresponds to 
 the birth of the protoplanetary disk, 
 which happens inside the outflow-launching region. 

The infalling from the envelope to the central region continues,  
 and the radius of the outflow-lauching region increases over time, 
 as does the outer boundary of the magnetic dead zone. 
Likewise, both the mass and the outer radius of the circumstellar disk 
 increase over time.  
Eventually, 
 the radius of the disk extends beyond the initial radius 
 of the first core and the first core disappears; 
in other words, the first core transforms itself into 
 a protoplanetary disk.

\begin{figure}
 \includegraphics[width=170mm]{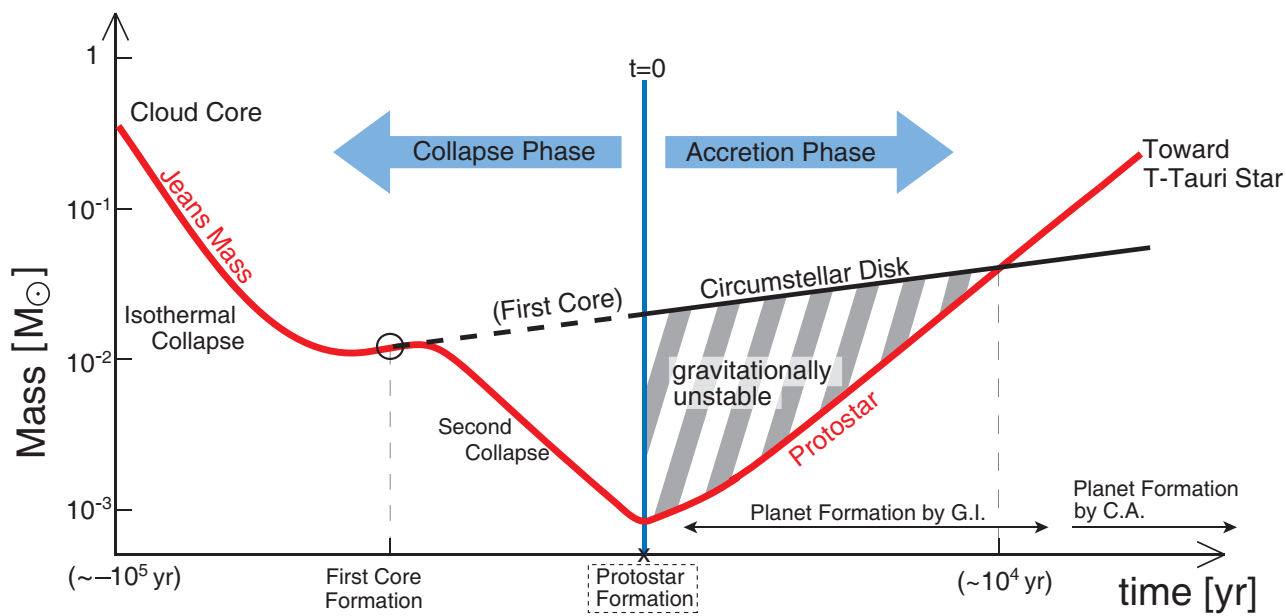}
\caption{
 \BLskip
 Schematic diagram for the evolution of protostellar objects, 
 in terms of mass. 
 The vertical axis denotes mass (in units of solar mass) and 
  the horizontal axis denotes time (in years). 
 The red curve on the left-hand side depicts the mass 
  of the fast collapsing region in the center of the molecular 
  cloud core in the collapsing phase, 
  which essentially defines the gravitationally 
  unstable mass, and therefore corresponds to the Jeans mass. 
Note that the mass of the first core is much larger than 
  the mass of the central protostar at its birth. 
The right-hand side describes the evolution in 
 the main accretion phase, where gas in the envelope of 
 the molecular cloud core accretes onto the central region and 
 the protostar gains its mass. 
As the first core gradually changes into the protoplanetary disk, 
 the mass of the protoplanetary disk remains larger 
 than the mass of the protostar for a while. 
This configuration is gravitationally unstable 
 and creates self-gravitating objects in the disk. 
The protostar mass increases monotonically and 
 overwhelms the mass of the disk later in the accretion phase.  
}
\label{fig:2}
\end{figure}
%
\begin{figure}[h]
 \includegraphics[width=170mm]{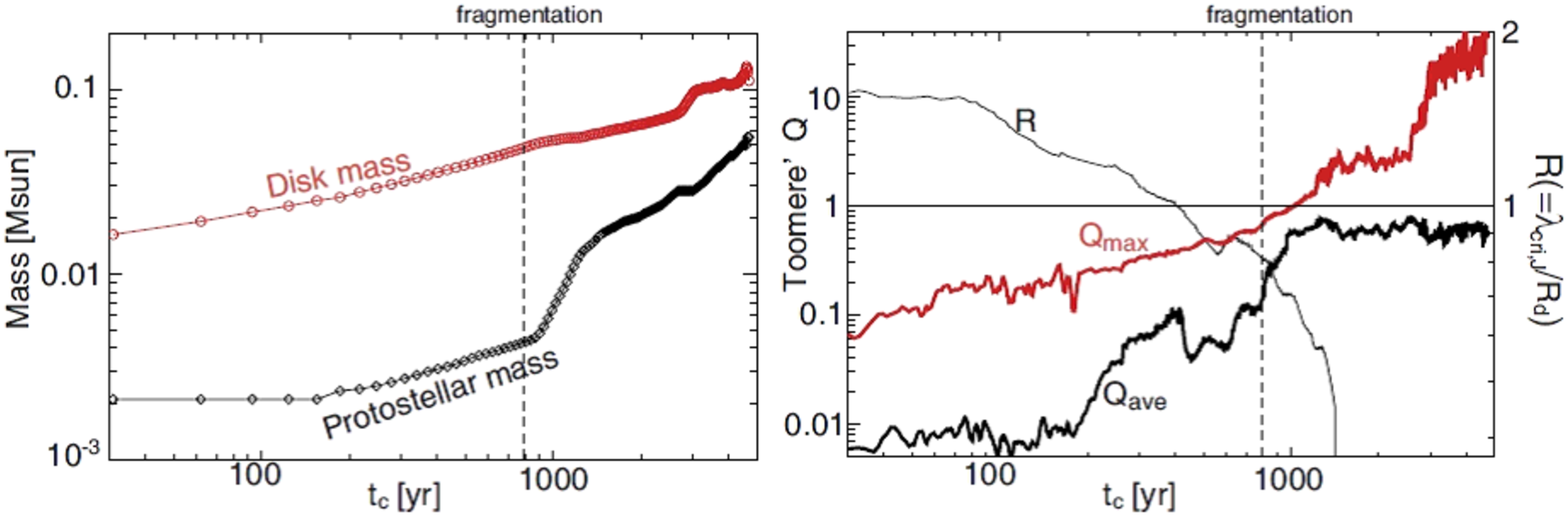}
\caption{
 \BLskip
Left panel shows actual evolution of protostar mass and disk mass, 
 where the horizontal axis denotes elapsed time in years after the formation 
 of the protostar. 
The right panel shows time evolution of 
 the gravitational instability indicator 
 $Q \equiv \kappa C_{\rm S}/(\pi G \Sigma)$. 
$Q_{\rm max}$ is the maximum value in the disk and 
 $Q_{\rm ave}$ is the average value. 
The curve labeled $R$ denotes the ratio of the most unstable 
 wavelength to the disk radius. 
At $t\approx~800$yr, $R$ becomes smaller than unity so that 
 the gravitationally unstable mode is possible in the disk, 
 and planetary-mass objects are formed. 
}
\label{fig:3}
\end{figure}

An important consequence of the early formation of 
 the protoplanetary disk is the gravitational fragmentation 
 of the disk and the formation of planetary-mass objects, 
 as shown in the lower right panel of Figure 1. 
The reason for the gravitational instability can be 
 understood in terms of the ratio of the disk mass to the central object. 
Figure 2 shows a schematic evolution of protostellar objects, 
 in terms of mass. 
The vertical axis denotes mass (in units of solar mass) and 
 the horizontal axis denotes time (in units of year). 
The red curve on the left-hand side depicts the mass 
 of the fast collapsing region in the center of the molecular 
 cloud core in the first collapse phase, 
 which essentially defines the gravitationally 
 unstable mass, and thus, corresponds to the Jeans mass. 
The decrease of this mass in the isothermal phase describes 
 the run-away collapse of central region, 
 a characteristic of gravitational collapse of cooling gas. 
The slight increase of the central mass (the Jeans mass) 
 corresponds to the formation of the first core and 
 its gradual increase in a quasi-steady state. 
The second fall of the central mass corresponds to 
 ``the second collapse'' that is triggered by 
 the endothermic dissociation of hydrogen molecules. 
After most of the hydrogen molecules are dissociated, 
 the adiabatic increase in the pressure of the atomic gas eventually overcomes 
 the gravitational collapse and the second core forms. 
Therefore, the decrease of the central mass in the second collapse stage 
 is closed by the formation of a protostar. 
We should note that the mass of the first core is much larger than 
 the mass of the central protostar at its birth. 

The right-hand side of Figure 2 describes the evolution in 
 the main accretion phase, where gas in the envelope of 
 the molecular cloud core accretes onto the central region and 
 the protostar gains mass. 
As the first core gradually changes into the protoplanetary disk, 
 the mass of the protoplanetary disk remains large (and even 
 larger than the protostar) for a while. 
This configuration is gravitationally unstable and the spiral arms 
 are excited, which promotes gas accretion in the disk, 
 but the accretion is not efficient enough to avoid 
 gravitational fragmentation of the disk, 
 creating self-gravitating objects in the disk. 
The typical mass of the formed objects is a fraction of the disk mass 
 (i.e., smaller than $10^{-2}$ solar masses) which corresponds to 
 the range of Jovian planets to brown dwarfs. 
The protostar mass increases monotonically and eventually 
 overwhelms the mass of the disk. 
In effect Figure 2 describes why and how the gravitationally 
 unstable protoplanetary disk is created.  

The left panel of Figure 3 shows the actual evolution 
 of the protostar mass and disk mass, where the horizontal 
 axis denotes elapsed time (in years after the formation 
 of the protostar). 
The right panel shows the time evolution of the indicator of 
 gravitational instability 
 $Q \equiv \kappa C_{\rm S}/(\pi G \Sigma)$, 
 where $\kappa$, $C_{\rm S}$, $G$, and $\Sigma$ are 
 epicyclic frequency, sound speed, gravitational constant, 
 and surface density, respectively. 
$Q_{\rm max}$ is the maximum value in the disk and 
 $Q_{\rm ave}$ is the average value. 
During the first several hundred years, $Q$ remains smaller than unity 
 and the disk satisfies the instability criterion 
 obtained by local linear analysis, 
 although the size of the disk remains too small for the unstable mode 
 to appear, as shown by the curve labeled $R$ that denotes 
 the ratio of the most unstable wavelength to the disk radius. 
At $t\approx~800$ yr, $R$ becomes smaller than unity so that 
 the gravitationally unstable mode becomes feasible in the disk;  
 this epoch does indeed correspond to the fragmentation of the disk. 
%
Note that the fragmentation has occurred 
 in the region where the effective equation of state is almost adiabatic. 
If additional radiative cooling operates, 
 it will become even more unstable, 
 and thus, the fragmentation is definitely expected. 
%
In other words, the gravitational fragmentation shown in this early 
 formation phase of the massive disk does not require efficient 
 radiative cooling 
 that is supposed to be important in the later phase of the evolution 
 where the disk is less-massive and less-unstable 
 \citep[e.g.,][]{Gammie2002,Rice+2003,RiceLodatoArmitage2005}. 
Here we emphasize that we should expect 
 the growth of gravitational instability in the disks that 
 excessively satisfy the local instability criterion ($Q~<~1$) 
 for the mode whose wavelength is sufficiently smaller than the disk. 
%
Nevertheless, the consequence of the growth and the resultant 
 fragmentation should be studied in more detail 
 including the effect of irradiation from the central star 
 that is not taken into account in the present calculations. 
%

The formation of planetary-mass objects and the excited spiral arm 
 structure provide an efficient transfer of angular momentum in the disk 
 and promote mass accretion onto the protostar, as shown by 
 the rapid increase in the protostar mass in the left panel. 
This property will be further investigated in our next paper. 

\section{Discussion}

\paragraph{Cooling Efficiency}
Since we are not solving realistic radiative transfer equation 
 in our dynamical simulations, one might wonder 
 whether each fragment can sufficiently shrink by radiating 
 an excess energy obtained by compressional heating or not. 
Here we estimate the amount of energy that should be radiated 
 away from a fragment and compare with the cooling capability 
 of the fragments. 
In order for the self-gravity to dominate over gas pressure 
 the effective ratio of specific heats ($\gamma_{\rm eff}$) 
 should be less than 4/3, while the ratio of specific heats 
 ($\gamma$) for the molecular gas in adiabatic evolution is mostly 7/5 
 \citep[e.g.,][]{MasunagaInutsuka2000}. 
Thus, we can roughly calculate the amount of energy $\Delta E$ 
 that should be radiated during the compressional motion 
 of the planet with mass $M_{\rm p}$
 from the density $\rho_{\rm d}$, 
 pressure $P_{\rm d}$, 
 and 
 temperature $T_{\rm d}$ 
 in the gas disk 
 to the average density $\rho_{\rm p}$ and pressure $P_{\rm p}$
 of the planet 
 as, 
\begin{equation}
 \Delta E = \frac{M_{\rm p}}{\gamma-1}
   \left( \frac{P_{\rm d}}{\rho_{\rm p}} \right)
   \left[ 
        \left(\frac{\rho_{\rm p}}{\rho_{\rm d}}\right)^{\gamma}
      - \left(\frac{\rho_{\rm p}}{\rho_{\rm d}}\right)^{\gamma_{\rm eff}}
   \right] . 
\end{equation}
If we take the density enhancement factor of $10^5$ 
 (e.g., $\rho_{\rm d} = 10^{-11} {\rm g~cm}^{-3}$ and 
        $\rho_{\rm p}    = 10^{-6} {\rm g~cm}^{-3}$)
 the second term is negligible compared to the first term 
 on the right-hand side of the above equation. 
Thus, the expected time-averaged luminosity of the fragment 
 that cools with the timescale $\Delta t$ 
 can be calculated as, 
\begin{equation}
  \langle L \rangle \equiv 
    \frac{\Delta E}{\Delta t} \approx  
    1.5 \times 10^{-1}
    \left( \frac{M_{\rm p}}{10^{-3} M_{\odot}} \right)
    \left( \frac{T_{\rm d}}{10^2{\rm K}} \right)
    \left( \frac{10^2 {\rm yr}}{\Delta t} \right)
    \left( \frac{\rho_{\rm p}}{10^5\rho_{\rm d}} \right)^{2/5}  
      L_{\odot} ,  
\end{equation}
where we assumed that the mean molecular weight is 2.3. 
On the other hand, the luminosity $L_{\rm p}$ of the planet with radius 
 $R_{\rm p}$ and the surface temperature $T_{\rm p}$ is 
\begin{equation}
  L_{\rm p} = 4 \pi R_{\rm p}^2 \sigma_{\rm SB} T_{\rm p}^4
    = 1.9 \times 10^{-1} 
      \left( \frac{R_{\rm p}}{10^{12}{\rm cm}} \right)^2
      \left( \frac{T_{\rm p}}{10^{3}{\rm K}} \right)^4
      L_{\odot} , 
\end{equation}
where $\sigma_{\rm SB}$ is Stephan-Boltzmann constant. 
Since $(T/T_{\rm d})=(\rho/\rho_{\rm d})^{\gamma_{\rm eff}-1}$, 
 compression by a factor of $\rho_{\rm p}/\rho_{\rm d} = 10^5$ 
 with $\gamma_{\rm eff}=4/3$ results in the increase of 
 temperature by a factor of about 46. 
Thus, the fragment can radiate efficiently enough to 
 collapse gravitationally by a factor of at least $10^5$ in density
 within 100 years  
 (i.e., the resultant surface temperature of the planet 
  will be somewhat smaller than $10^3$K). 
Thus, the objects formed by the gravitational instability 
 of the disk as in our simulation are expected to be dense enough 
 to avoid being washed out in the dynamical environment.

\paragraph{Further Evolution}
What is the fate of these gravitationally formed planetary-mass objects? 
Does gravitational interaction with the gaseous disk lead to 
 the  migration of these objects? 
In fact, some of our simulations showed the objects falling into 
 the central star, while in other cases they migrated outward 
 but remained in the disk. 
Their chaotic behavior is not surprising, since eccentricities of the 
 orbits of planetary-mass objects in the disk are, in general, not small. 
Their large eccentricities are due to the fact that they absorb 
 infalling gas. 

In addition, the gravitational fragmentation and formation of 
 planetary-mass objects repeat many times during the main accretion phase, 
 where the gas accretion from the envelope of the molecular 
 cloud core continues.  Some of the planetary-mass objects 
 fall onto the central star while some remain in the disk. 
Therefore, a statistical method seems to be required to predict 
 the distribution of objects that survive the gas dispersal, 
 which will be the scope of our subsequent paper. 

\paragraph{Implication for Observations}
Astronomical observation of the early formation phase of protoplanetary 
 disks would enable us to directly test these expectation. 
Although our numerical simulations enable us to frequently observe 
 the gravitational fragmentation of protoplanetary disks, 
 in reality this process is hidden in dense gas and may remain 
 invisible in optical and near-infrared wavelengths. 
However, radio interferometers are potentially capable of observing 
 the interiors of protostellar cloud cores. 
The Atacama Large Millimeter Array (ALMA) would be an ideal tool for 
 this type of observation.  
In order to directly observe the process described in this Letter, 
 it is necessary to observe the early evolutionary phase within about 
 1000 years from the birth of a protostar. 
As the average lifetime of a protostar is estimated to be $10^5$ year,  
 the required time span for this observation is only 1 \% of 
 the protostar's lifetime, meaning that 
 we should observe, on average, about 100 protostars to find one protostar 
 at an early evolutionary phase. 
In our simulations, we almost always observe very time-dependent 
 mass accretion through the gravitationally unstable disk. 
Therefore, another possible observational signature is 
 the time variability of the protostellar source. 
We hope that the gravitational fragmentation reported in this Letter 
 will be analyzed observationally and a realistic planet formation scenario 
 will be refined in the near future. 

The frequent formation of planetary-mass objects in the disks during the 
 formation phase suggests that we should consider their influence 
 on the evolution of the protoplanetary disk. 
In particular, 
 it suggests a new possibility for the core-accretion scenario 
 in the region outside the orbit of the planetary-mass object 
 where the shepherding effect from a massive object prevents 
 dust grains from raining out onto the central star 
 \citep{MutoInutsuka2009}, 
 which corresponds to the far right-side of Figure 2.  
Therefore, it might be interesting to search for a hybrid scenario of 
 planet formation 
 where the rocky planets form after the gravitational formation of 
 the gaseous giant planets.


This work was supported in part by Grants-in-Aid for Scientific Research 
from the MEXT of Japan 
(15740118, 16077202, 18540238, 18740104, 20540238, and 21740136). 



\end{document}